\documentclass[aps]{revtex4}
\usepackage{color}
\usepackage{graphicx}
\usepackage{eurosym}
\usepackage{graphicx}
\usepackage{amsmath,amssymb}
\usepackage{color}
\usepackage[outdir=./]{epstopdf}
\usepackage{amsmath}
\usepackage{amsfonts}
\usepackage{amssymb}

\def\bc{\begin{center}}
\def\ec{\end{center}}

\def\beq{\begin{eqnarray}}
\def\eeq{\end{eqnarray}}

\def\bc{\begin{center}}
\def\ec{\end{center}}

\def\beq{\begin{eqnarray}}
\def\eeq{\end{eqnarray}}

\begin{document}

\title{
Solar sail with inflatable toroidal shell }
\author{ V. Ya. Kezerashvili$^{1}$, R. Ya. Kezerashvili$^{1,2}$, O. L.
Starinova$^{3}$}
\affiliation{\mbox{$^{1}$Physics Department, New York City College
of Technology, The City University of New York,} \\
Brooklyn, NY 11201, USA \\
\mbox{$^{2}$The Graduate School and University Center, The
City University of New York, \\
New York, NY 10016, USA }\\
\mbox{$^{3}$Samara National Research University, Russian Federation, Samara,
Russia }\\
}
\date{\today }

\begin{abstract}
In the framework of a strict mathematical approach based on classical theory
of elasticity we present an idea of the deployment and stretching of the
circular solar sail attached to the inflatable toroidal shell. It is predicted
that by introducing the gas into the inflatable toroidal shell one can
deploy and stretch a large size circular solar sail membrane. The formulas
for the toroidal shell and sail membrane stresses and strains caused by the
gas pressure in the shell are derived. The analytical expressions can be
applied to a wide range of solar sail sizes. Numerical calculations for the
sail of radii up to 100 m made of CP1
membrane and attached to the toroidal shell with the varied cross-section
radius are presented. %We study the deflection of the torus-shaped sail flat membrane due to the acceleration by solar radiation pressure.
The normal transverse vibration modes of the sail membrane under tension caused by gas pressure in the shell are calculated.
The feasibility of deployment and stretching of a solar sail with a large size circular membrane attached to the inflatable toroidal shell is demonstrated.

\end{abstract}

\maketitle

\section{Introduction}

Currently the main space exploration vehicle relies on chemical propulsion
system and most missions provide energy by means of prelaunch onboard fuel.
A solar sail is propellant-less propulsion system for space
exploration that uses the Sun radiation
as a propulsion mechanism. A solar sail is a large sheet of low areal
density material that gains an acceleration due to the reflection and
absorption of the Sun electromagnetic flux \cite{pol, Colin, Matloff3,
MatloffValpetti}. The proposals to use solar sails cover almost the whole
spectrum of space missions, from an expedition to Mars to scientific probes,
from continuous planetary polar observation to mining exploitation of the
asteroids, even implementations for deep space exploration and interstellar
travel are considered. The successful deployment of the world's first
interplanetary solar sail IKAROS, NASA's first solar NanoSail-D and the
Interplanetary Society solar sail LightSail-2 demonstrates the feasibility of
exploration of the Solar system by means of the electromagnetic radiation
pressure.

Studies on a solar sail have four primary foci: i. finding low areal density
material that allows to utilize the maximum acceleration due to the solar
radiation pressure; ii. missions design for exploration of the Solar System
and beyond using the solar sail; iii. maximizing  the solar thrust through the
increase of area of the solar sail made of a low areal density material; iv.
the development of the mechanism for the deployment and stretching the large
size of solar sail membrane. In this work we concentrated on the last two
objectives.

The sail's membrane deployment strategies has attracted considerable
attention. Many different systems have been previously considered for the
sails opening. The deployment is usually performed: by uniaxial mechanisms,
such as the  telescopic, deployable and inflatable booms; the extendable masts; the centrifugal force that renders a spin-type deployment
mechanism; by the presence of guide rollers; electromechanical actuation
devices, or composite booms; the solar sail self-deployment based on shape
memory alloys,  (see Refs. \cite{Genta1999,9,Deployment4,8,91,MemorySail,
MemorySail2,Deployment1,Deployment2,Deployment3,Dep4,Dep5} and references therein) and most recently the deployment of the large size solar sail was considered using the superconducting current-carrying wire \cite{VKRK20221,VKRK20222}. We
cite these works, but the recent literature on the subject is not limited by
them. It is worth mentioning that in the actual deployment technology, the
main limit is still the high weight of the system and the complexity of the
deployment mechanism for solar sail surface.

The concept of using inflatable structures for a spacecraft has been
extensively discussed during the past six decades. In the 1960s NASA
launched the Echo balloons that were some of the first inflatable structures
to be utilized in space \cite{5}. Once such a structure is inflated and
deployed, it operates by similar to rigid structures principles and allows to
obtain comparable or greater performance. The advantages of inflatable
structures are a simple deployment mechanism for large space structures, the
small mass and space savings in the launch configuration \cite{4}.

Inflatable structures have the characteristics that are particularly
advantageous for a solar sail: i. extremely lightweight that is critical for
a solar sail; ii. a simple mechanism to deploy it in orbit.

The use of inflatable structures for solar sailing goes back to 1989 when
Strobl \cite{Strobl1} proposed a hydrogen inflated hollow disk-shaped sail
with a molybdenum reflector. Later the idea of the inflatable ring sail was
presented in Ref. \cite{Hayn}. Inflatable structures including the solar
sail have been the subject of interest in the past and recent years and
investigated %in detail
in Refs. \cite%
{Strobl2,Genta1999,Matloff1,Kezmetloff1,Kezmetloff2,8,Leigh,
Tinker,KezASR2021}. Most recently a sail consisting of a
reflective membrane attached to an inflatable torus-shaped shell was suggested
\cite{KezASR2021}. The sail deployment from its stowed configuration is
initiated by introduction of the inflation pressure into the toroidal shell.
However, the study \cite{KezASR2021}, as well as the previous investigations did not address important issues related to the elastic properties
of the sail membrane and torus-shaped shell. The toroidal shell is kept pressurized by the
gas after deployment, withholds all the stresses, and allows a very simple straightforward deployment procedure of a large
area circular configuration.

In this work we consider the toroidal shell that is attached to a circular
solar sail membrane to aid the membrane deployment and stretching. The
sail's deployment from its stowed configuration is initiated by introducing
inflation pressure into the toroidal shell. We provide the rigorous
consideration of this system within the framework of the classical theory of
elasticity \cite{Timoshenko51,Landau7}. An important question that arises in
the context of deployable toroidal-shaped solar sail is its mass and
stability. We consider statics of a deployed torus-shaped sail, and address the structural
strength of the sail membrane and inflated toroidal shell to support the
flat surface of a circular membrane.
%The deflection of the flat membrane due to accelerations initiated by the solar radiation pressure is studied.
The governing equation for a gas in an inflated toroidal shell is assumed to be
the equation for a real gas. Within such approach we investigate the effects
of both the enclosed gas pressure and structure stiffness and the stability of the solar sail. It is
demonstrated that the effect of the enclosed gas must be considered in the
analysis of the inflatable torus with the membrane.

This article is organized in the following way. In Sec. \ref{theory} we consider the solar sail configuration and within the framework of the
classical theory of elasticity derive the
forces that lead to the deployment of a circular sail
membrane attached to the toroidal shell. The stress and strain of the toroidal shell, the circular
membrane
%under the uniformly distributed force applied to the membrane edge as well as the stress and strain in the toroidal
and the toroidal shell-membrane structure
are addressed. % within the theory of elasticity.
%The diflection of the sail membrane due to the solar radiation pressure and acceleration of the sail as well as
The vibration of the deployed and stretched membrane are considered in Sec. \ref{membrane}.
Results of the numerical calculations and discussion are
presented in Sec. \ref{results}. The conclusions follow in Sec. \ref%
{conclusions}.

\section{Solar Sail as a Circular Membrane Attached to a Toroidal Shell }

\label{theory}

A schematic of the circular solar sail attached to a torus of radius $R$ and
a circular cross-section of radius $r$ inflated by a gas is presented in
Fig. \ref{Torusfig1}. The radius of the toroidal shell is significantly
greater than the radius cross-section: $R\gg r$. The solar sail membrane is
stretched by forces generated by pressure of the gas introduced into the
toroidal shell. In this Section we consider the strain in a toroidal shell
due to the gas pressure and provide the detailed consideration of the stress
and strain in the circular membrane resulting from uniformly distributed
force applied to the membrane edge and caused by the gas pressure.
Finally, the system of the toroidal shell filled by the gas
attached to the sail's circular membrane is considered.

\subsection{Toroidal shell filled with gas}

\begin{figure}[h]
\noindent
\begin{centering}
\includegraphics[width=7.2cm]{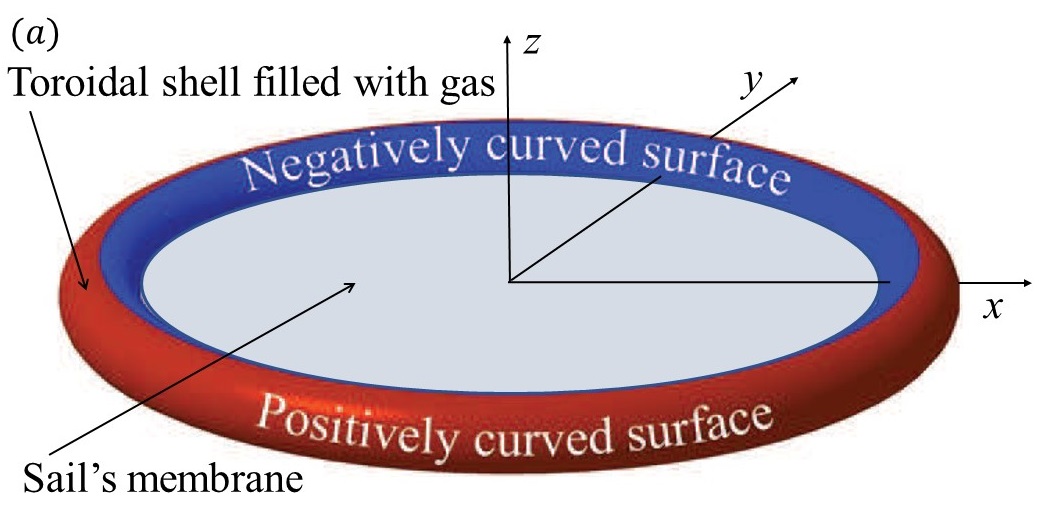}
\includegraphics[width=7.8cm]{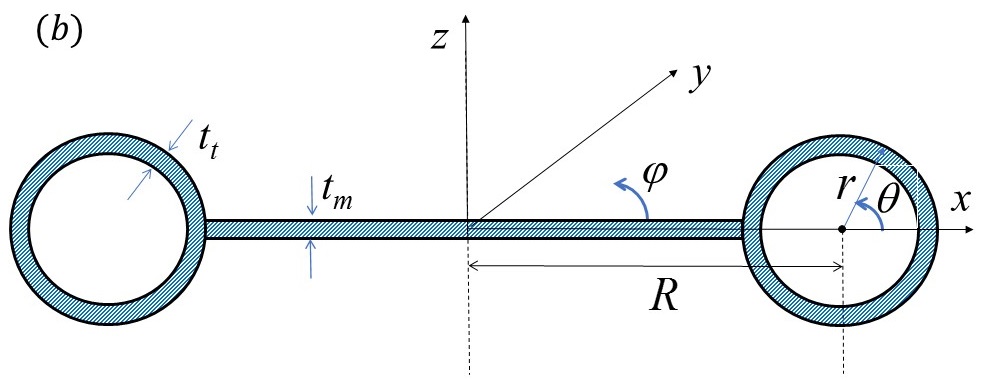}
\par\end{centering}
\caption{(Color online) ($a$) Torus-shaped solar sail and its geometric
characteristics. ($b$) $R$ and $r$ are the radii of the torus and the toroidal
shell cross-section, respectively. Throughout of this work we assume that
the radius of the torus is much greater than the radius of the toroidal
shell: $R\gg r$. $t_{m}$ and $t_{t}$ are the thicknesses of the sail
membrane and toroidal shell, respectively.}
\label{Torusfig1}
\end{figure}

A key component and basic structural element of the torus-shaped solar sail
is a toroidal shell. The gas filled toroidal shell attached to the
circular membrane (Fig. \ref{Torusfig1}$a$) provides structural support to the sail. Here we consider
a static behavior of an inflated toroidal shell following Ref. \cite%
{Kraus1967}. We assume large aspect ratio, $\frac{R}{r}\gg 1$. Therefore, $%
R\gg r$ and it is a very reasonable approximation to consider that the
radius of the circular sail membrane $R-r\simeq R$ is the radius of the
torus. Also the thickness of the torus shell $t_{t}$ is negligible compared
to the radii of curvature.

An implicit equation for a torus with a large aspect ratio\ that is
azimuthally symmetric about the $z-$axis in Cartesian coordinates is
\begin{equation}
(R-\sqrt{x^{2}+y^{2}})^{2}+z^{2}=r^{2},  \label{torus1}
\end{equation}%
where $R$ is the distance from the center of the toroidal shell to the
center of the torus and $r$ is the radius of the toroidal shell as is shown
in Fig. \ref{Torusfig1}b. This torus is generated by revolving the circle $%
(x-R)^{2}+z^{2}=r^{2}$ of radius $r$ in the $x-z$ $-$ plane about the $z$
axis at the distance $R$ from the center of the circle.

In our consideration it is convenient to use toroidal and poloidal
coordinates for description of the torus geometry: the polar coordinate $%
(r,\theta )$ for each cross-section and toroidal coordinate $\varphi $
around the toroidal ring. The three coordinate $r,\theta ,\varphi $ ($\theta
\in \left[ 0,2\pi \right] ,$ $\varphi \in \left[ 0,2\pi \right] $) of curved
coordinate system are orthogonal to each other. The toroidal coordinate is
used to describe those quantities along the direction normal to the
cross-section, while the poloidal coordinates are used to describe an
interior of the toroidal shell. The toroidal and poloidal coordinate system
relates to standard Cartesian coordinates as follows
\begin{eqnarray}
x &=&(R+r\cos \theta )\cos \varphi ,  \notag \\
y &=&(R+r\cos \theta )\sin \varphi ,  \label{Torcoordinate} \\
z &=&r\sin \theta .  \notag
\end{eqnarray}
\begin{figure}[t]
\centering
\includegraphics[width=6.1cm]{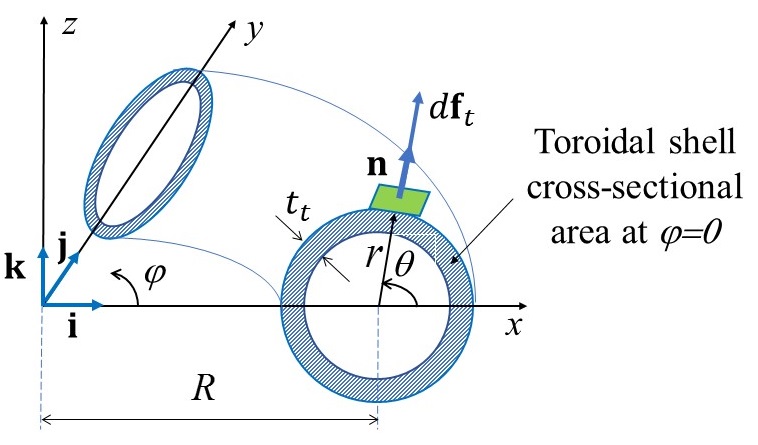}
\caption{(Color online) A schematic of the element of the toroidal shell
inflated by the gas. The force $df_{t}$ acts on the surface element of the
shell at the angle $\protect\theta $ per unit of the length of toroidal ring
at the angle $\protect\varphi =0$. The radius $R$ is much greater
than the cross-section radius of the torus $r$: $R\gg r$. The figure is not
to scale.}
\label{Fig1a}
\end{figure}
Of our special interest are the forces that act on the surface area of the torus.
The total surface area of the torus can be considered as a sum of the outer
and inner surfaces as shown in Fig. \ref{Torusfig1}$a$. The outer surface is
a positively curved in a contrast of the inner surface, which is negatively
curved. It is obvious that the outer surface is larger than the inner one
and the difference in surface area is $\Delta A=8\pi r^{2}$. Therefore, the pressure of the
gas $P$ in toroidal shell filled with the gas produces the larger force on
the outer surface of the shell than on the inner one.
\begin{figure}[t]
\centering
\includegraphics[width=9.0cm]{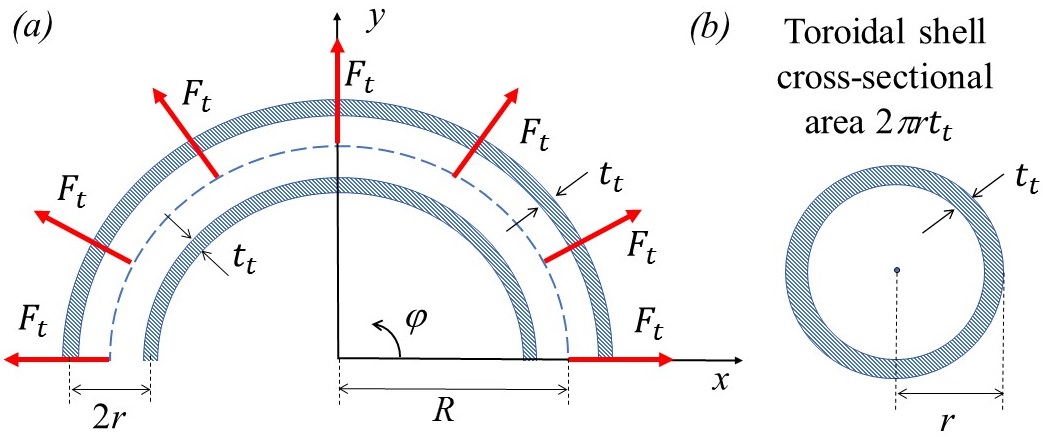}
\caption{(Color online) ($a$) Schematics for the action of uniform radial forces $%
F_{t}$ per unit of circumferential length on the toroidal shell.
($b$) The area of the toroidal shell cross-section is $A = 2\protect\pi r t_{t}$.
The figure is not to scale: $R\gg r$ and $r\gg t_{t}$}
\label{Fig5}
\end{figure}
Let us find the net force due to the gas pressure $P$ in the toroidal
shell that keeps the torus inflated. We consider the torus circular
cross-section at the angle $\varphi =0$ as it is depicted in Fig. \ref{Fig1a}%
. The differential of the force $d\mathbf{f}_{t}$ acting on the surface
element $dA$ at angle $\theta $ per unit of the length of toroidal ring $%
Rd\varphi $\ is $d\mathbf{f_{t}=}\frac{PdA\widehat{\mathbf{n}}}{Rd\varphi }$,
where $dA=(R+r\cos \theta )d\varphi rd\theta $ and $\widehat{\mathbf{n}}%
=\cos \theta \mathbf{i+\sin }\theta \mathbf{k}$ is a unit vector normal to
the surface element $dA$ with the unit vectors $\mathbf{i}$ and $\mathbf{k}$
along $x$ and $y$ axis, respectively. Therefore, the net force that acting
on the unit length of the toroidal shell at $\varphi =0$ is
\begin{equation}
\mathbf{F}_{t}=\int_{0}^{2\pi }\frac{P}{R}(R+r\cos \theta )\left( \cos
\theta \mathbf{i+\sin }\theta \mathbf{k}\right) rd\theta =\frac{\pi Pr^{2}}{R%
}\mathbf{i}.  \label{Total}
\end{equation}%
The force $\mathbf{F}_{t}$ is directed along $x$ axis radially out of the
torus center, which is expected due to the mirror symmetry of the torus with
respect of the $x-y$ plane. The magnitude of the force $\mathbf{F}_{t}$ at $%
\varphi =0$ acting on the unit length of the toroidal shell is
\begin{equation}
F_{t}=\frac{\pi Pr^{2}}{R}.  \label{Wtotal}
\end{equation}%
Due to the fact that the $z-$axis is the axis of rotational symmetry of
infinite order the magnitude of the force $F_{t}$ for any toroidal angle $%
\varphi \in \left[ 0,2\pi \right] $ is the same as (\ref{Wtotal}) for $%
\varphi =0$ and is directed radially out as shown in Fig. \ref{Fig5}. This
force is responsible for the tensile stress $\sigma _{t}$ of the toroidal
shell resulting in tensile strain $\Delta R_{t}/R$.

The net force acting on the toroidal shell cross-section of the aria $2\pi
rt_{t}$ is $\frac{1}{2}\int_{0}^{\pi }F_{t}\sin \varphi Rd\varphi =F_{t}R$.
In the limit of the linear response one can apply the Hook's law and obtains
the tensile stress in the toroidal shell \cite{Timoshenko51}
\begin{equation}
\sigma _{t}=\frac{F_{t}R}{2\pi rt_{t}}=E_{t}\frac{\Delta R_{t}}{R}.
\label{WR}
\end{equation}%
In Eq. (\ref{WR}) %and (\ref{DeltaRR})
$E_{t}$ and $t_{t}$ are Young's
modulus and the thickness of the toroidal shell material, respectively.
%In Eqs. (\ref{WR}) and (\ref{DeltaRR}) $E_{t}$ and $t_{t}$ are Young's
%modulus and the thickness of the toroidal shell material, respectively.
From Eq. (\ref{WR}) %and (\ref{Wtotal})
we get for the strain and $\Delta R_{t}$ the
following expression:
\begin{equation}
\frac{\Delta R_{t}}{R}=\frac{F_{t}R}{2\pi E_{t}rt_{t}},  \label{DeltaRR}
\end{equation}%
\begin{equation}
\Delta R_{t}=\frac{F_{t}R^{2}}{2\pi E_{t}rt_{t}}.  \label{Sarea}
\end{equation}%
This $\Delta R_{t}$ is for a stand-alone torus. In the equilibrium of the
stand-alone gas filled torus under the force $F_{t}$ is balanced by the
opposite elastic forces.

The toroidal shell
is made from a material with density $\rho _{t}$ and the density of
the material of the membrane  %which has the area $A=\pi R^{2},$
is $\rho_{m} $.
The thicknesses of the foil for the toroidal shell and membrane
are $t_{t}$ and $t_{m}$, respectively. Thus, the mass of the sail is
\begin{equation}
M=\pi \rho _{m}R^{2}t_{m} + 4\pi ^{2}\rho _{t}rRt_{t}+M_{g},
\label{Mass}
\end{equation}%
where ${M_{g}}$ is the mass of the gas.

\subsection{Stress and strain of circular membrane}

\begin{figure}[t]
\centering
\includegraphics[width=11.0cm]{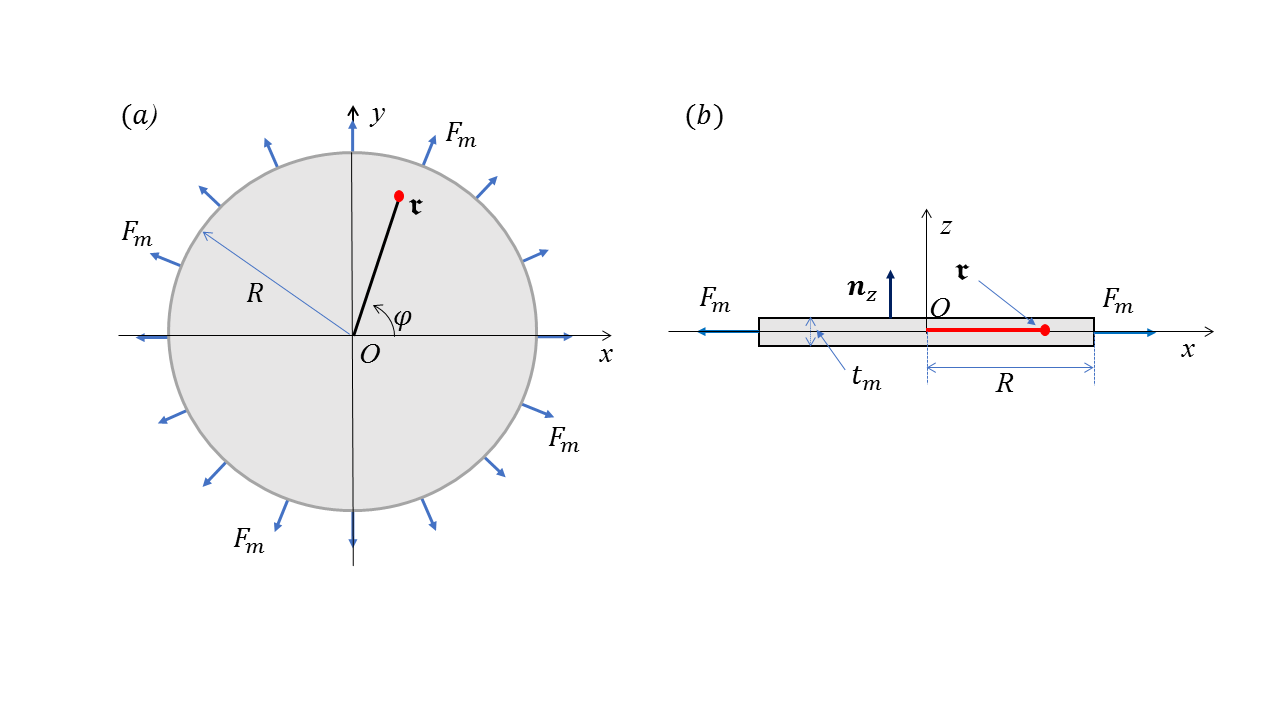}
\caption{(Color online) Schematics for the action of the uniformly
distributed radial force $F_{m}$ per unit of circumferential length of the
circular membrane: ($a$) top view; ($b$) side view. The figure is not to
scale: $R\gg t_{m}$ }
\label{Fig6}
\end{figure}

Ideally the torus-shaped solar sail has the attached membrane of the radius $%
R-r$. However, as we mentioned before for the torus-shaped sail with large
aspect ratio when $R\gg r$, the radius of the membrane can be considered as $%
R $. Consider a thin circular membrane of the radius $R$ and thickness
$t_{m}$ ($t_{m}\ll R$) under a uniform distributed force $F_{m}$ per unit
length of a circumference acting at the membrane edge in the radial
direction as shown in Fig. \ref{Fig6}. If the membrane is sufficiently thin,
the deformation can be treated as uniform over its thickness and we have to
deal with longitudinal deformations of the membrane and not with any
membrane bending. For a two-dimensional case the strain tensor is a function
of $x$ and $y$ coordinates and is independent on $z$. The boundary
conditions for the stress tensor on both surfaces of the membrane are \cite%
{Landau7}
\begin{equation}
\sigma _{ik}n_{k}=0,n_{k}=n_{z},i=x,y,z,
\end{equation}%
where $\mathbf{n}_{z}$ is the normal vector parallel to $z-$axis, and lead
to
\begin{equation}
\sigma _{xz}=\sigma _{yz}=\sigma _{zz}=0
\end{equation}%
in the entire volume of the membrane when $t_{m}\ll R$ \cite{Landau7}.

The equation of equilibrium in the absence of the body forces in the
two-dimensional vector form is \cite{Landau7}:
\begin{equation}
grad\text{ }div\text{ }\mathbf{u-}\frac{1}{2}(1-\nu _{m}) curl\text{ }curl\text{ }\mathbf{u}=0,  \label{Curl}
\end{equation}%
where $\mathbf{u}$ is the displacement vector, $\nu _{m}$ is the membrane
Poisson ratio and the vector operators $grad$, $div$ and $curl$ are two-dimensional. Due to the
axial symmetry $\mathbf{u}$ is directed along the radius and is a function
of $\mathfrak{r}$ only, so $curl$ $\mathbf{u}=0$ and Eq. (\ref{Curl})
in polar coordinates becomes:

\begin{equation}
div\text{ }\mathbf{u=}\frac{1}{\mathfrak{r}}\frac{d(\mathfrak{r}u)}{d%
\mathfrak{r}}=const\equiv 2c
\end{equation}%
and gives
\begin{equation}
u=c\mathfrak{r}+\frac{d}{\mathfrak{r}}.  \label{usolutionG}
\end{equation}%
Therefore, the stress tensor radial and angular components are
\begin{equation}
u_{\mathfrak{rr}}=\frac{du}{d\mathfrak{r}}=c-\frac{d}{\mathfrak{r}^{2}},%
\text{ }u_{\varphi \varphi }=\frac{u}{\mathfrak{r}}=c+\frac{d}{\mathfrak{r}%
^{2}},  \label{usolution}
\end{equation}%
where in (\ref{usolution}) $c$ and $d$ are some constants \cite{Handbook}.
In polar coordinates at $\varphi =0$ the stress $\sigma _{ik}$ and strain $%
u_{ik}$ tensor components are:

\begin{eqnarray}
\sigma _{\mathfrak{rr}} &\mathbf{=}&\sigma _{xx},\text{ }\sigma _{\varphi
\varphi }\mathbf{=}\sigma _{yy},\text{ }  \label{sigsig} \\
u_{\mathfrak{rr}} &\mathbf{=}&u_{xx},\text{ }u_{\varphi \varphi }\mathbf{=}%
u_{yy}.  \label{uu}
\end{eqnarray}%
The general equations relating the strain tensor components to the stress
tensor components \cite{Landau7} in our case when $\sigma _{zz}=0$ become:
\begin{eqnarray}
\sigma _{xx} &=&\frac{E_{m}}{1-\nu _{m}^{2}}\left( u_{xx}+\nu
_{m}u_{yy}\right) ,  \label{Sigxx} \\
\sigma _{yy} &=&\frac{E_{m}}{1-\nu _{m}^{2}}\left( u_{yy}+\nu
_{m}u_{xx}\right) ,  \label{Sigyy}
\end{eqnarray}%
where $E_{m}$ and $\nu _{m}$ are the Young modulus of elasticity and Poisson
ratio of the membrane, respectively. By substituting (\ref{sigsig}) and (\ref%
{uu}) at $\varphi =0$ into (\ref{Sigxx}) and (\ref{Sigyy}) the stress tensor
radial and angular components become:
\begin{eqnarray}
\sigma _{\mathfrak{rr}} &\mathbf{=}&\frac{E_{m}}{1-\nu _{m}^{2}}\left( u_{%
\mathfrak{rr}}+\nu _{m}u_{\varphi \varphi }\right),  \label{sigmarr} \\
\sigma _{\varphi \varphi } &\mathbf{=}&\frac{E_{m}}{1-\nu _{m}^{2}}\left(
u_{\varphi \varphi }+\nu _{m}u_{\mathfrak{rr}}\right).  \label{sigmarr2}
\end{eqnarray}%
The requirement for the deformation $u$ (\ref{usolution}) to be finite at
the membrane center and the boundary condition for $\sigma _{\mathfrak{rr}}$
(\ref{sigmarr}) combined with $u_{\mathfrak{rr}}$ and $u_{\varphi \varphi }$
(\ref{usolution}) at the membrane edge:

\begin{equation}
u\left( 0\right) =0\text{ and }\sigma _{\mathfrak{rr}}(\mathfrak{r}=R)=\frac{%
F_{m}}{t_{m}}  \label{bondary}
\end{equation}%
determine the values of the constants $c$ and $d$:%
\begin{equation}
c=\frac{F_{m}}{t_{m}}\frac{(1-\nu _{m})}{E_{m}},\text{ }d=0.  \label{const}
\end{equation}%
The use of the constants (\ref{const}) allows to obtain the expressions for the
deformation vector $\mathbf{u}$ (\ref{usolutionG}) and the strain (\ref%
{usolution}), as well as the stress tensor
components (\ref{sigmarr}), (\ref{sigmarr2}):%
\begin{eqnarray}
\mathbf{u} &\mathbf{=}&\frac{F_{m}}{t_{m}}\frac{(1-\nu _{m})}{E_{m}}%
\mathfrak{r}, \\
u_{\mathfrak{rr}} &=&u_{\varphi \varphi }=\frac{F_{m}}{t_{m}}\frac{(1-\nu
_{m})}{E_{m}},  \label{Strdist} \\
\sigma _{\mathfrak{rr}} &=&\sigma _{\varphi \varphi }=\frac{F_{m}}{t_{m}}.
\label{St45}
\end{eqnarray}%
As expected the stress distribution (\ref{St45}) in the membrane deformed by
the forces acting at the edge of the membrane does not depend on the
elasticity constants of the membrane media \cite{Landau7}. Finally, both the
radial deformation $\Delta R_{m}=u(\mathfrak{r}=R)$, the strain $\frac{%
\Delta R_{m}}{R}(\mathfrak{r}=R)$ and the stress $\sigma _{\mathfrak{r}%
m}=\sigma _{\mathfrak{rr}}(\mathfrak{r}=R)$ of the membrane edge are:
\begin{eqnarray}
\Delta R_{m} &\mathbf{=}&\frac{F_{m}}{t_{m}}\frac{(1-\nu _{m})}{E_{m}}R,
\label{Dbmembrane} \\
\frac{\Delta R_{m}}{R} &=&\frac{F_{m}}{t_{m}}\frac{(1-\nu _{m})}{E_{m}},
\label{Dbmembrane1} \\
\sigma _{\mathfrak{r}m} &=&\frac{F_{m}}{t_{m}}.  \label{Dbmembrane2}
\end{eqnarray}%
It is important to mention that all expressions in SubSection II.B are valid for a membrane of thickness $t_{m}$ in the
shape of a ring with concentric internal and external
radii, subject to internal membrane clamping. This is still
the same case of two-dimensional uniform expansion as
considered above.
\begin{figure}[t]
\centering
\includegraphics[width=8.0cm]{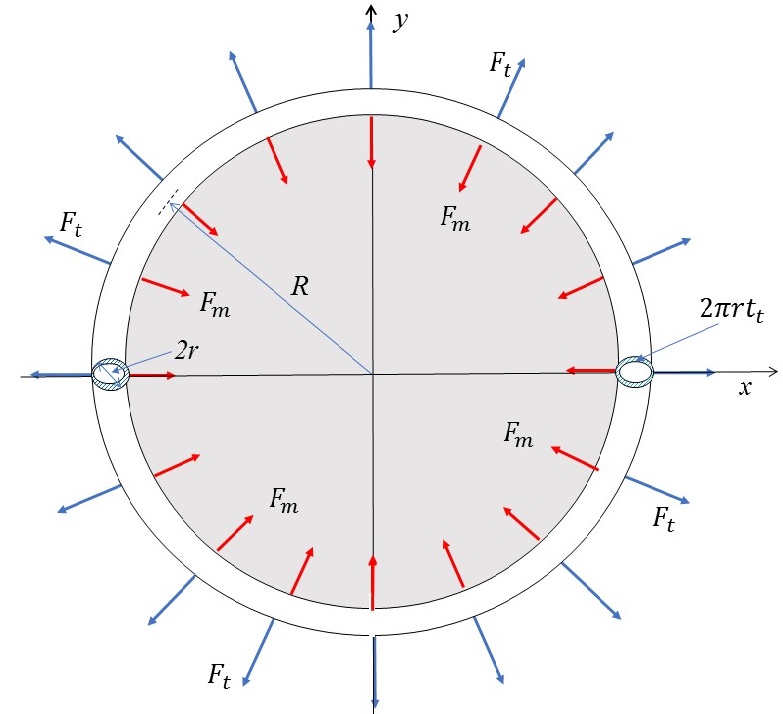}
\caption{(Color online) Schematics for the action of uniform radial forces $%
F_{t}$ and $F_{m}$ per unit of circumferential length on the toroidal shell.
The force opposite in direction and equal in magnitude to $F_{m}$ acts on
the circular membrane and is not shown. For visibility, the cross-section of
the toroidal shell is given in 3D format. The figure is not to scale: $R\gg
r $ and $r\gg t_{t}$.}
\label{Fig51}
\end{figure}

\subsection{Gas in toroidal shell}

Consider the pressure of the gas in the toroidal shell. We
describe the gas by the equation for the real gas. Therefore, we are not
considering atoms or molecules of the gas as point particles that interact
with toroidal shell only. We are taking into account the volume that a real
gas atoms or molecules take up and gas atoms or molecules interaction with
each other that are experiencing an attraction at low pressure and repulsion at
high pressure. The corresponding van der Waals equation of state reads
\begin{equation}
\left( P+a\frac{n^{2}}{V^{2}}\right) \left( V-nb\right) =nR_{g}T.
\label{Klaiperon}
\end{equation}%
In Eq. (\ref{Klaiperon}) $a$\ and $b$\ are the gas dependent constants, $n=%
\frac{M_{g}}{\mu }$ is the number of moles of the gas, where $M_{g}$ is the
mass of the gas not considering diffusion losses and $\mu $ is the molar
weight of the gas, $V$ and $T$ are volume and temperature of the gas,
respectively, and $R_{g}=8.31$ J$\cdot $K$^{-1}$mol$^{-1}$is a universal gas
constant. We assume that at equilibrium the pressure of the gas in the
toroidal shell obeys to the isochoric process because it is confined in the
interior volume $V=2\pi ^{2}r^{2}R$,\ which only slightly changes because of
the increase of the torus radius by $\Delta R_{t}$ (\ref{Sarea}) due to the
gas pressure. For the torus with large aspect ratio this increase of the
volume can be neglected. The internal pressure produces the forces that
remain normal to the surface. As it follows from Eq. (\ref{Klaiperon}) for
the isochoric process the pressure
\begin{equation}
P=n\frac{R_{g}T}{2\pi ^{2}r^{2}R-nb}-n^{2}\frac{a}{4\pi ^{4}r^{4}R^{2}}.
\label{pressure}
\end{equation}%
As it follows from Eq. (\ref{pressure}) the pressure of the confined gas is
defined by its mass, temperature, the gas parameters $a$\ and $b$, and the characteristic size of the torus radii
$R$ and $r$.

Using Eqs. (\ref{Wtotal}) \ and (\ref{pressure}) one can find the in-plane
tensile force per unit length in the membrane produced by the inflatable
torus shell filled with a gas
\begin{equation}
F_{t}=n\frac{\pi r^{2}R_{g}T}{2\pi ^{2}r^{2}R^{2}-nbR}-n^{2}\frac{a}{4\pi
^{3}r^{2}R^{3}}.  \label{TencsilForce}
\end{equation}%

\subsection{Circular membrane attached to toroidal shell}

In the absence of the gas, there are no stresses along the line of the
attachment of the membrane to the toroidal shell. The gas in the toroidal
shell causes the force $F_{t}$ acting on the toroidal shell radially out and
$F_{m}$ acting on the toroidal shell from the membrane radially in as shown
in Fig. \ref{Fig51}. The latter force is in return to the stress caused by
the force of the toroidal shell on the membrane equal in magnitude
to $F_{m}$. In the state of equilibrium we have

\begin{equation}
\Delta R_{t}=\Delta R_{m}.  \label{rav}
\end{equation}%
The radial forces $F_{t}$ and $F_{m}$ result in total radial force $%
F_{t}-F_{m}$, which has to be balanced by the elastic forces in the toroidal
shell causing the shell radius change $\Delta R_{t}$ according to (\ref%
{Sarea}). However, the force $F_{t}$ in (\ref{Sarea}) for stand-alone toroidal shell should be replaced
by the force $F_{t}-F_{m}$ for the toroidal shell and membrane combination.
Using (\ref{Sarea}) in place of $\Delta R_{t}$ with $F_{t}$ replaced to $%
F_{t}-F_{m}$, and using (\ref{Dbmembrane}) for the $\Delta R_{m}$ of
membrane leads (\ref{rav}) to
\begin{equation}
\frac{(F_{t}-F_{m})R^{2}}{2\pi rt_{t}E_{s}}=\frac{F_{m}}{t_{m}}\frac{(1-\nu
_{m})}{E_{m}}R,
\end{equation}%
and
\begin{equation}
F_{m}=F_{t}\left( 1+2\pi \frac{r}{R}\frac{t_{t}}{t_{m}}\frac{E_{t}}{E_{m}}%
(1-\nu _{m})\right) ^{-1}.  \label{wm}
\end{equation}%
Using (\ref{Wtotal}) for $F_{t}$ Eq. (\ref{wm}) can be rewritten as
\begin{equation}
F_{m}=\pi P\frac{r^{2}}{R}\left( 1+2\pi \frac{r}{R}\frac{t_{t}}{t_{m}}\frac{%
E_{t}}{E_{m}}(1-\nu _{m})\right) ^{-1},
\label{Fm}
\end{equation}%
where $P$ is defined by (\ref{pressure}).

In conclusion, (\ref{Dbmembrane1}) with (\ref{Fm}) for $F_{m}$ %and (\ref{TencsilForce}) for $F_{t}$
allow the determination of the strain of both
the toroidal shell and the membrane. Equation (\ref{Dbmembrane2}) determines
the radial pressure on the membrane edge, and (\ref{WR}) with $F_{t}$
replaced by $F_{t}-F_{m}$:
\begin{equation}
\sigma _{t}=\frac{\left( F_{t}-F_{m}\right) R}{2\pi rt_{t}}
\label{StressM}
\end{equation}%
provides for the tensile stress of the toroidal shell for the system of the
circular membrane attached to the toroidal shell.

\section{Sail membrane}

\label{membrane} The sail surface is a circular membrane of radius $R$
represents the "still" drum head shape under the tensile stress produced
by the inflated toroidal shell. The boundary of the membrane is a circle of
radius $R$ centered at the origin and represents the rigid frame to which
the membrane is attached. In the equilibrium position, the membrane is
stretched and fixed along its entire boundary by the toroidal shell in the $%
x-y $ plane. The tension per unit length $F_{m}$ caused by stretching the
membrane is the same at all points and in all directions and does not change
during the motion.

When the deployment of the membrane ends it can experience the vibration.
The mathematical equation that governs the normal transverse vibrations of the membrane is the
wave equation with zero boundary conditions. The wave equation has been
widely studied in the literature and is as follows:
\begin{equation}
\frac{\partial ^{2}\mathfrak{u}}{\partial t^{2}}=c_{s}^{2}\left( \frac{%
\partial ^{2}\mathfrak{u}}{\partial x^{2}}+\frac{\partial ^{2}\mathfrak{u}}{%
\partial y^{2}}\right) ,\text{ \ }c_{s}^{2}=\frac{F_{m}}{\varkappa },
\label{waveeq}
\end{equation}%
where $\mathfrak{u}(x,y)=0$ on the boundary of the membrane, $\varkappa $ is
the mass per area of the membrane,\ and $c_{s}=\sqrt{\frac{F_{m}}{%
\varkappa }}$ is the sound speed. Equation (\ref{waveeq}) is the
two-dimensional wave equation, which is a second order partial differential
equation. Due to the circular geometry of the membrane, it is convenient to
rewrite Eq. (\ref{waveeq}) using the cylindrical coordinates
\begin{equation}
\frac{\partial ^{2}\mathfrak{u}}{\partial t^{2}}=c_{s}^{2}\left( \frac{%
\partial ^{2}\mathfrak{u}}{\partial \mathfrak{r}^{2}}+\frac{1}{\mathfrak{r}}%
\frac{\partial \mathfrak{u}}{\partial \mathfrak{r}}+\frac{1}{\mathfrak{r}^{2}%
}\frac{\partial \mathfrak{u}}{\partial \varphi }\right) ,\text{ \ for }0\leq
\mathfrak{r}<R,\text{ and \ }0\leq \varphi \leq 2\pi  \label{Radial}
\end{equation}%
with the boundary condition $\mathfrak{u}(\mathfrak{r},\varphi ,t)=0$ for $%
\mathfrak{r}=R,$ which means that the membrane is fixed along the boundary
circle of the radius $R$ in the $x-y-$plane for all times $t\geq 0$.

To determine solutions $\mathfrak{u}(\mathfrak{r},\varphi ,t)$ that are
radially symmetric, we solve Eq. (\ref{Radial}) following the three standard
steps: i. using the method of separation of variables, we first determine
solutions as $\mathfrak{u}(\mathfrak{r},\varphi ,t)=\mathfrak{U}(\mathfrak{r}%
,\varphi )V(t)$ and obtain two independent differential equations for $%
\mathfrak{U}(\mathfrak{r},\varphi )$ and $V(t)$ functions; ii. from the
solutions of those ordinary differential equations we determine solution
(eigenfunctions) $\mathfrak{U}(r,\varphi )$ which satisfy the boundary
condition $\mathfrak{U}(\mathfrak{r},\varphi )=0$ for $\mathfrak{r}=R$ and
find the corresponding eigenvalues. Then find the periodical solution $V(t)$
by solving the second differential equation; iii. we compose these solutions
and obtain the radially symmetric solution $\mathfrak{u}(\mathfrak{r},t)$ that satisfy
the conditions $\mathfrak{u}(\mathfrak{r},0)$ and $\frac{\partial \mathfrak{u}(\mathfrak{r},t)}{%
\partial t}$ depend only on $\mathfrak{r}.$ The corresponding method of a solution is
given in Ref. \cite{KezASR2021}. In-plane vibrations and dumping of the transfer vibrations in particular due to the solar radiation pressure are out of the score of this study.

\section{Results and discussion}

\label{results} The theory presented in the previous sections provides
expressions to calculate different quantities related to a torus-shaped
solar sail such as stresses in the toroidal shell and membrane, the radial
force that acts on the membrane and their dependence on the gas mass. The
general assumptions of applicability of these expressions are the following:

i. both strain and stress for the membrane and toroidal shell are small and
within the limit of elasticity;

ii. the radius of the membrane is much greater than the radius of the
toroidal shell: $R\gg r$. %Therefore, the aspect ratio $\frac{R}{r}\gg 1$;

iii. the material thickness of the membrane and toroidal shell are significantly
smaller than the radius of the toroidal shell: $t_{m}\ll r$ and $t_{t}\ll r$.

The first condition insures the mechanical stability of the sail and requires that the stress in the membrane and toroidal shell
have to be less than the yield strength of the corresponding material. To
satisfy the second condition the values of $\frac{R}{r}\sim 10^{2}$ are
used. To meet the third requirement the values $\frac{t_{m}}{r}%
\approx \frac{t_{t}}{r}\sim 10^{-5}$ are considered.
\begin{table}[b]
\caption{Model parameters for the sail membrane and hydrogen gas. $\rho _{m}$,
$E_{m}$, $\nu _{m}$ and  $t_{m}$ are the density, Young
modulus, Poisson ratio and thickness of the sail membrane,
respectively.}
\label{tab1}
\begin{center}
\begin{tabular}{cccc|ccc}
\hline\hline
\multicolumn{4}{c}{Sail's membrane, CP1} & \multicolumn{3}{|c}{Gas parameters
} \\ \hline
$\rho _{m}$, kg/m$^{2}$ & $E_{m}$, Pa & $\nu _{m}$ & $t_{m}$, m & $\mu ,$ g/mol
& $a$, L$^{2}$bar/mol$^{2}$ & $b,$ L/mol \\ \hline
1.43$\times 10^{3}$ & 2.17$\times 10^{9}$ & 3.40$\times 10^{-1}$ & 3.5$%
\times 10^{-6}$ & 2.0159 & 0.2476 & 0.02661 \\ \hline
\end{tabular}
\end{center}
\end{table}
In our calculations we consider CP1 Polyimide film \cite{CP1} as an example of the material
of the membrane and toroidal shell (CP1 is planned to be used for the Solar Cruiser \cite{LesJ}) and the hydrogen as the filling gas. However, to meet the first condition the
thickness of the toroidal shell is considered twofold as for the membrane.
As it follows from Eqs. (\ref{Wtotal}), (\ref{wm}), and (\ref{StressM}) this leads almost twofold decrease of the shell stress. The properties of CP1, corresponding parameters and gas constants are listed in Table \ref%
{tab1}. The yield strength of CP1 is not available to the authors.
%Although for the real design considerations the yield strength has to be measured experimentally,
In general, it is much less than the Young modulus. Nonetheless, one
can estimate the yield strength to be greater than $\sim $10$^{-2}$ part of
the Young modulus, keeping in mind that the real design decisions have to be
based on experimentally measured values.

\begin{figure}[h]
\noindent
\begin{centering}
\includegraphics[width=6.2cm]{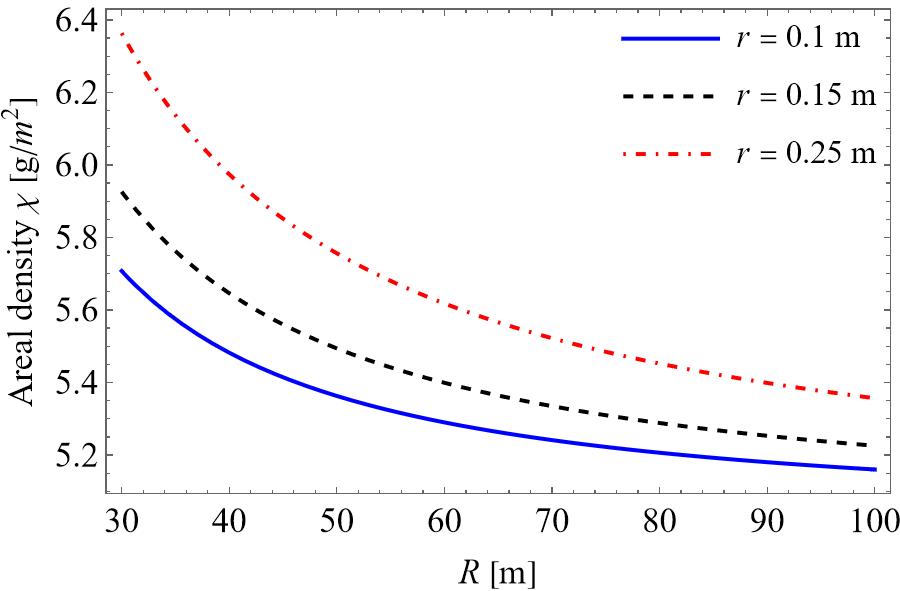}
\par\end{centering}
\caption{(Color online) The dependence of the areal density the solar sail
on the membrane radius for different radii of the toroidal shell.}
\label{ArealDensity}
\end{figure}

\begin{figure}[h]
\noindent
\begin{centering}
\includegraphics[width=6.2cm]{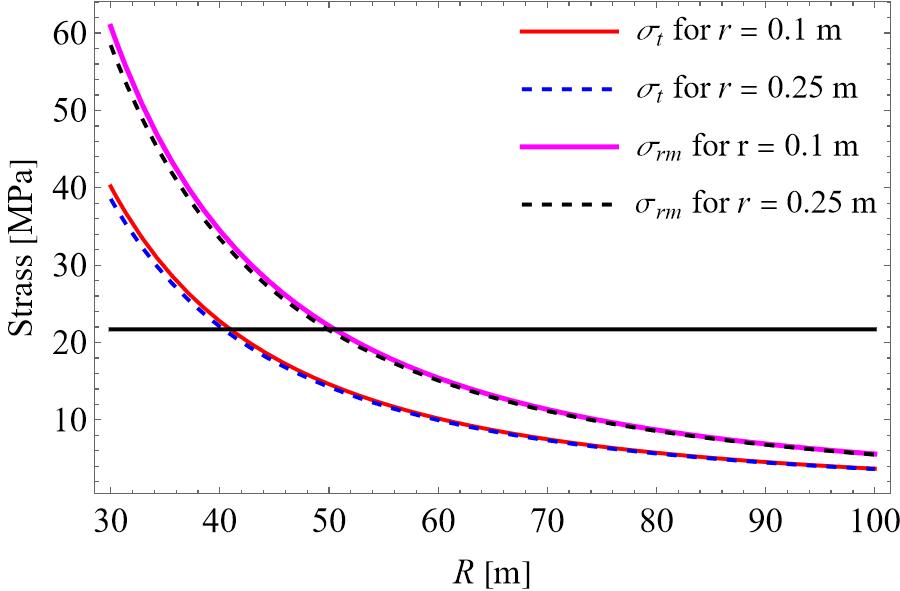}
\par\end{centering}
\caption{(Color online) The dependence of the tensile stress of the toroidal
shell $\sigma_{t}$ and the membrane radial stress $\sigma _{\mathfrak{r}m}$ on the membrane radius for different radii of the toroidal
shell. The horizontal line corresponds to 10$^{-2}E_{m}$ MPa. Calculations performed for the gas mass $M_{g} = 0.750$ kg. }
\label{Stresses}
\end{figure}

\begin{figure}[h]
\noindent
\begin{centering}
\includegraphics[width=6.2cm]{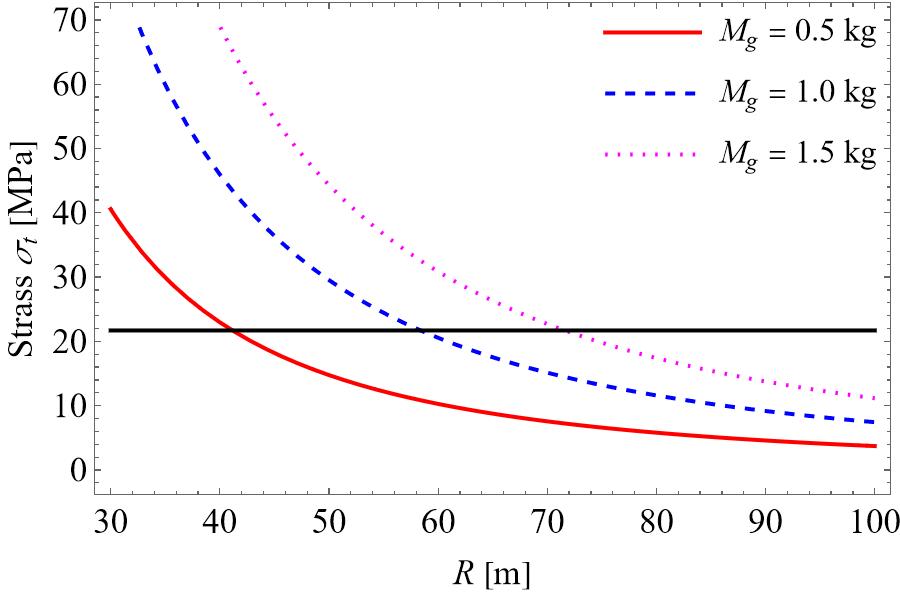}
\par\end{centering}
\caption{(Color online) The dependence of the membrane radial stress $\sigma _{\mathfrak{r}m}$ %forces $F_{m}$
on the
membrane radius for different masses of the gas in the toroidal shell. The horizontal line corresponds to 10$^{-2}E_{m}$ MPa. Calculations performed for $r = 0.1$ m. }
\label{ForceFmm}
\end{figure}

\begin{figure}[h]
\noindent
\begin{centering}
\includegraphics[width=6.2cm]{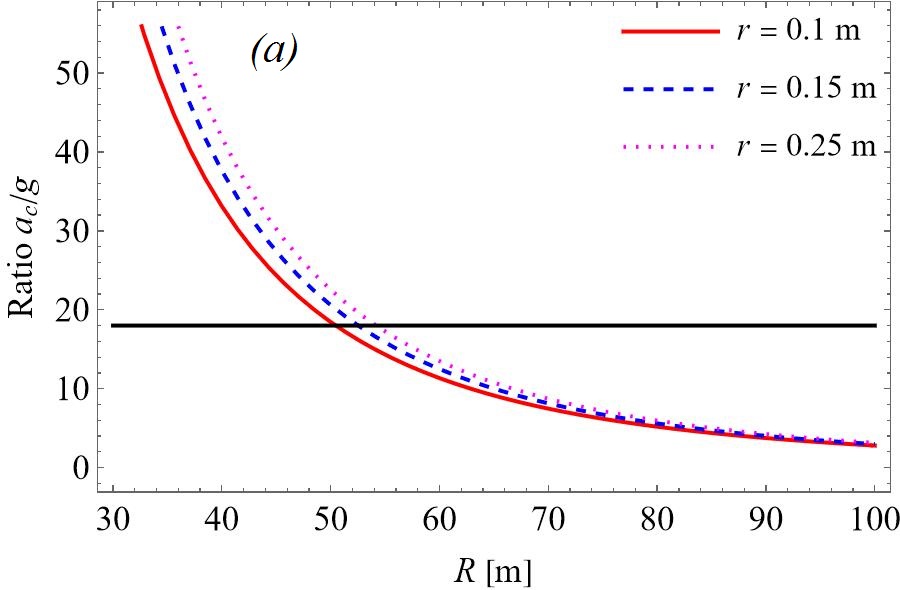}
\includegraphics[width=6.2cm]{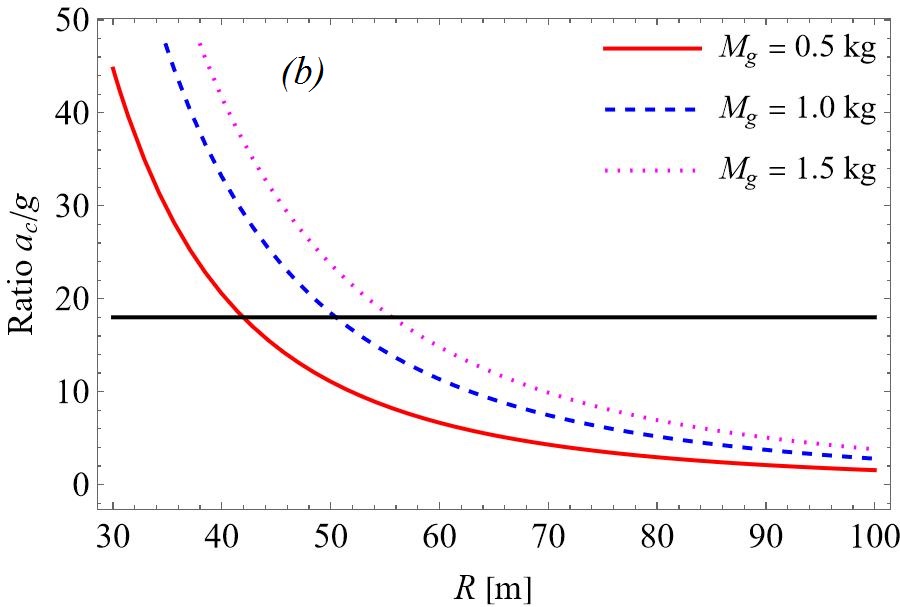}
\par\end{centering}
\caption{(Color online) The dependence of the ratio $a_{c}/g$ on the radius
of the solar sail membrane for ($a$) different radii of the toroidal shell for $M_{g}$ = 1.0 kg
and ($b$) different masses of the gas in the toroidal shell for $r = 0.1$ m. The horizontal
line for the ratio $a_{c}/g$ is obtained based on the result from Ref.
\protect\cite{Deployment4}.}
\label{Ratio}
\end{figure}
\begin{figure}[h]
\noindent
\begin{centering}
\includegraphics[width=5.2cm]{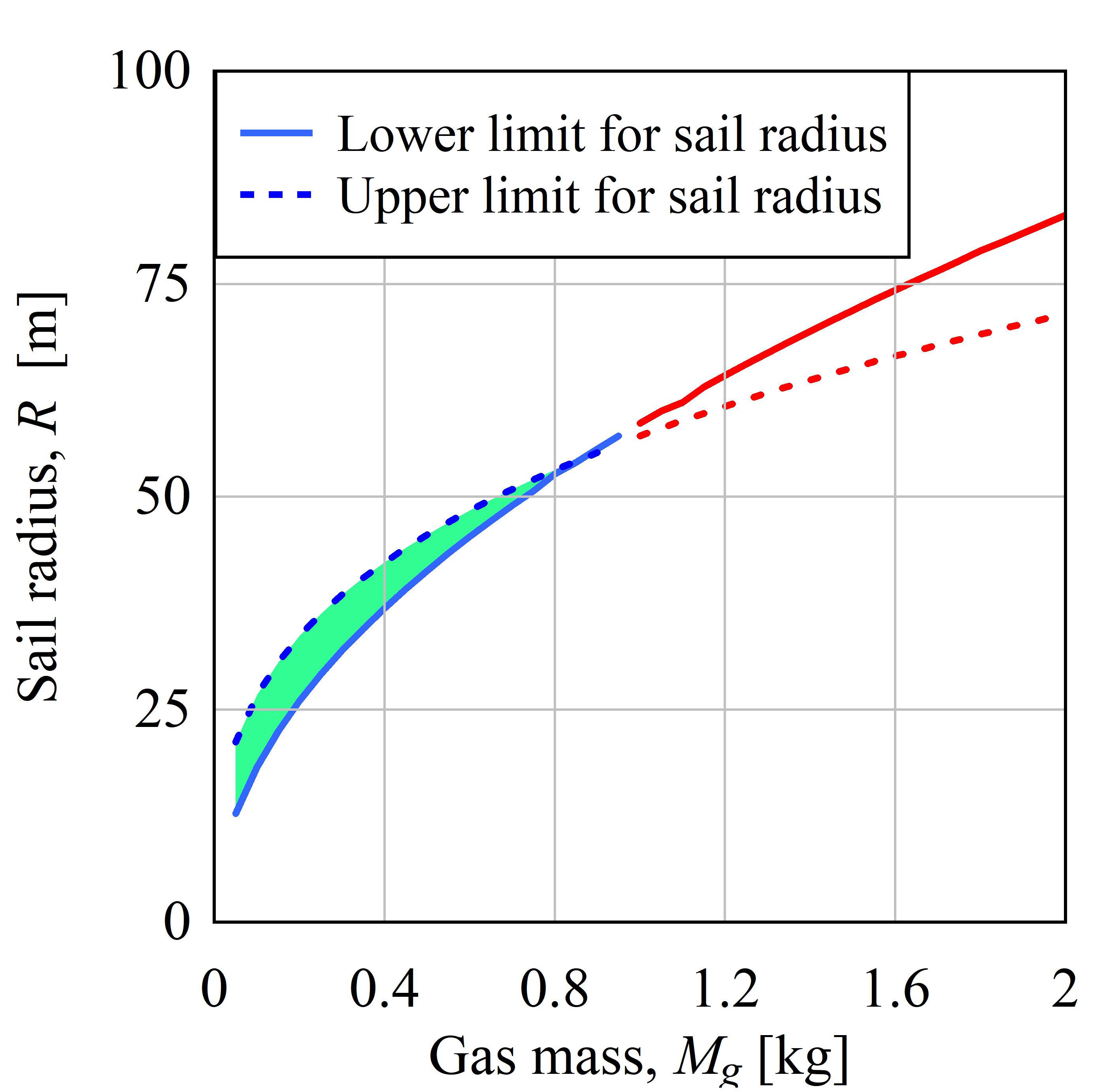}
\par\end{centering}
\caption{(Color online) Limitations for the radius of the sail. The colored area shows the allowed radii of the sail's membrane for the radius of the toroidal shell $r =0.1$ m.
}
\label{Lim}
\end{figure}
One of the key metrics related to the performance of solar sail is the
characteristic acceleration \cite{Colin,Matloff3,MatloffValpetti}. The
characteristic acceleration is defined as $a_{0}=2\eta P_{0}/\chi $, where $%
0.5\leq \eta \leq 1$, $P_{0}$ is the solar radiation pressure near the Earth
and $\chi =M/A$ is the areal density. In the latter expression, $M$ is the
total mass of the torus-shaped solar sail defined by Eq. (\ref{Mass}) and $A$ is the sail membrane area. It is worth mentioning that we neglect the projectional area of the toroidal shell: $4\pi r(R+r) << \pi R{^2}$. However, consideration of this area only decreases the areal density. While the actual sail
acceleration is a function of heliocentric distance and its orientation, the
characteristic acceleration allows a comparison of solar sail design
concepts on an equal footing. The areal density $\chi $ is an important
parameter determining the sail acceleration due to electromagnetic radiation pressure. It is
easy the calculate $\chi $ for known geometry and densities of sail
materials. Less $\chi $ leads to the greater sail acceleration. The dependence of the areal density of the solar sail on the membrane radius for different
radii of the toroidal shell is shown in Fig. \ref{ArealDensity}. As is expected the areal density increases with
the increase of the toroidal shell radius and decrease of the sail membrane radius. When the membrane radius increases from 30 m to 100 m the areal density decreases form  $\sim$ 6.4 g/m$^{2}$ to $\sim$ 5.2 g/m$^{2}$. While the dependence of $\chi $ on the toroidal shell radius is important, the mass of the gas has negligible contribution.

The expressions (\ref{Dbmembrane2}) and (\ref{StressM}) are used to calculate the radial stress of the membrane and tensile stress of the toroidal shell. Analysis of these expressions shows the strong dependence on the membrane radius and on the pressure of the gas, but negligible dependence on the radius of the toroidal shell. Results of calculations performed for the different fixed radii of the toroidal shell are presented in Fig. \ref{Stresses}. The both stresses decrease with the increase of membrane radius and the results show that for any given radius the membrane stress is greater than respective toroidal shell stress. Thus, the stability of the membrane will insure the stability of the entire shell-membrane structure. The membrane stress stays within 10$^{-2}E_{m}$.
%and satisfy the first condition: both the toroidal shell and membrane stresses stay within the 10$^{-2}$ range of the corresponding Young modulus  %Thus, they are within the limit of elasticity.
%for the radii greater than the crossover point radii of respective stress versus radius curves with the horizontal line drawn at the $10^{-2}$ of Young modulus level.
The crossover points radii represent the lower limits of the sail radii.
% From the results presented in Fig. \ref{Stresses} it follows that for the given radius of the membraine the membrane stresses are always greater than respective toroidal shell stresses. Thus, the stability of the membrane will insure the stability of the entire toroidal shell-membrane structure. The toroidal shell-membrane sails need to have the radii greater than radii of the crossover points for respective  membrane stresses to remain within the limits of elasticity and stability.

In Eq. (\ref{pressure}) the second term is small and one can say that the gas pressure in the toroidal shell is approximately directly proportional to the gas mass. Therefore, analysis of Eqs. (\ref{TencsilForce}) and (\ref{wm}) shows that the force $F_{t}$ per unit length of the toroidal shell and the force $F_{m}$ per unit length of the membrane %produced by the torus shell strongly depends on the gas mass and
also are  approximately directly proportional to the gas mass.
%In Fig. \ref{ForceFmm} is shown the dependence of the radial forces $F_{m}$ on the membrane radius for different masses of
%the gas in the toroidal shell. The force $F_{t}$ is a little bit bigger than $F_{m}$ by the factor
%$\left( 1+2\pi \frac{r}{R}\frac{t_{t}}{t_{m}}\frac{%
%E_{t}}{E_{m}}(1-\nu _{m})\right)$ and has the same kind of dependence on $R$ and the gas mass as $F_{m}$.  For the large radius of the sail membrane the dependence on the mass significantly decreases.
The same strong dependence on the gas mass along with the strong dependence on the membrane radius is reflected on both the toroidal shell and membrane stresses. The results presented in Fig. \ref{ForceFmm} show the dependence of the membrane stress on the membrane radius for different masses of the gas. The strong dependence of the radii of the crossover points on the gas mass is obvious. The stable sails need to have radii greater than the radii of crossover points in Fig. \ref{ForceFmm}.

The logical question which arises now is the following: which value of the
sail radius should be considered to deploy the sail successfully? Although a certain answer can be
obtained only in the experiment it is possible to make some estimations by introducing the ratio $a_{c}/g$ given as
\begin{equation}
\frac{a_{c}}{g}=\frac{\left( F_{t}-F_{m}\right) }{g(2\pi r t_{t}\rho _{t} + M_{g}/2\pi R)},
\label{ratio}
\end{equation}%
where %$\rho _{t}$ is the toroidal shell mass density and
$g=9.8$ m/s$^{2}$ is the
acceleration due to gravity at the Earth surface. The quantity $a_{c}$ is
effective acceleration equivalent to centripetal acceleration of the
shell would it be spinning around its center with frequency
\begin{equation}
f=\frac{1}{2\pi }\left( \frac{a_{c}}{R}\right) ^{1/2}.
\end{equation}%
The greater $a_{c}/g$ ratio the greater the chance of successful deployment
of initially folded toroidal shell and sail membrane to the open state
of circular shape. A simple analysis of the partial derivatives shows that $%
\sigma _{t}$, $\sigma _{rm}$, and $a_{c}/g$ are the monotonically decreasing
functions of the membrane radius $R$, radius $r$ of the toroidal shell cross-section and increasing functions of the gas mass.

We compare the calculated
values of $a_{c}/g$ ratio with the equivalent estimate based on the
successful ground demonstration of the spinning sail deployment by Jet
Propulsion Laboratory \cite{Deployment4}. The data provided in Ref. \cite%
{Deployment4} for the spinning frequency 200 RPM or more
of the sail of 0.8 m in diameter gives the estimate of the ratio of
centripetal acceleration over the acceleration due to gravity equal to 18 or
more. The calculated $a_{c}/g$ ratio is presented in Fig. \ref{Ratio}. It is evident from Fig. \ref{Ratio}$a$ that the dependence of the ratio on the toroidal shell radius $r$ is fairy weak and similar to the dependence of the stresses on $r$ - see Fig. \ref{Stresses}. Thus, the choice of lower $r$ is preferable for the sail due to the less arial density. Our choice for toroidal radius $r$ is 0.1 m. Results presented in Fig. \ref{Ratio}$b$ show strong dependence of the $a_{c}/g$ ratio on the gas mass. The radii of the crossover points of the $a_{c}/g$ versus $R$ curves with the horizontal line at $a_{c}/g$ = 18 represent the upper limits of the sail radius. The sail with radius less than these upper limits is expected to successfully deploy into a circular shape.
\begin{figure}[b]
\noindent
\begin{centering}
\includegraphics[width=5.2cm]{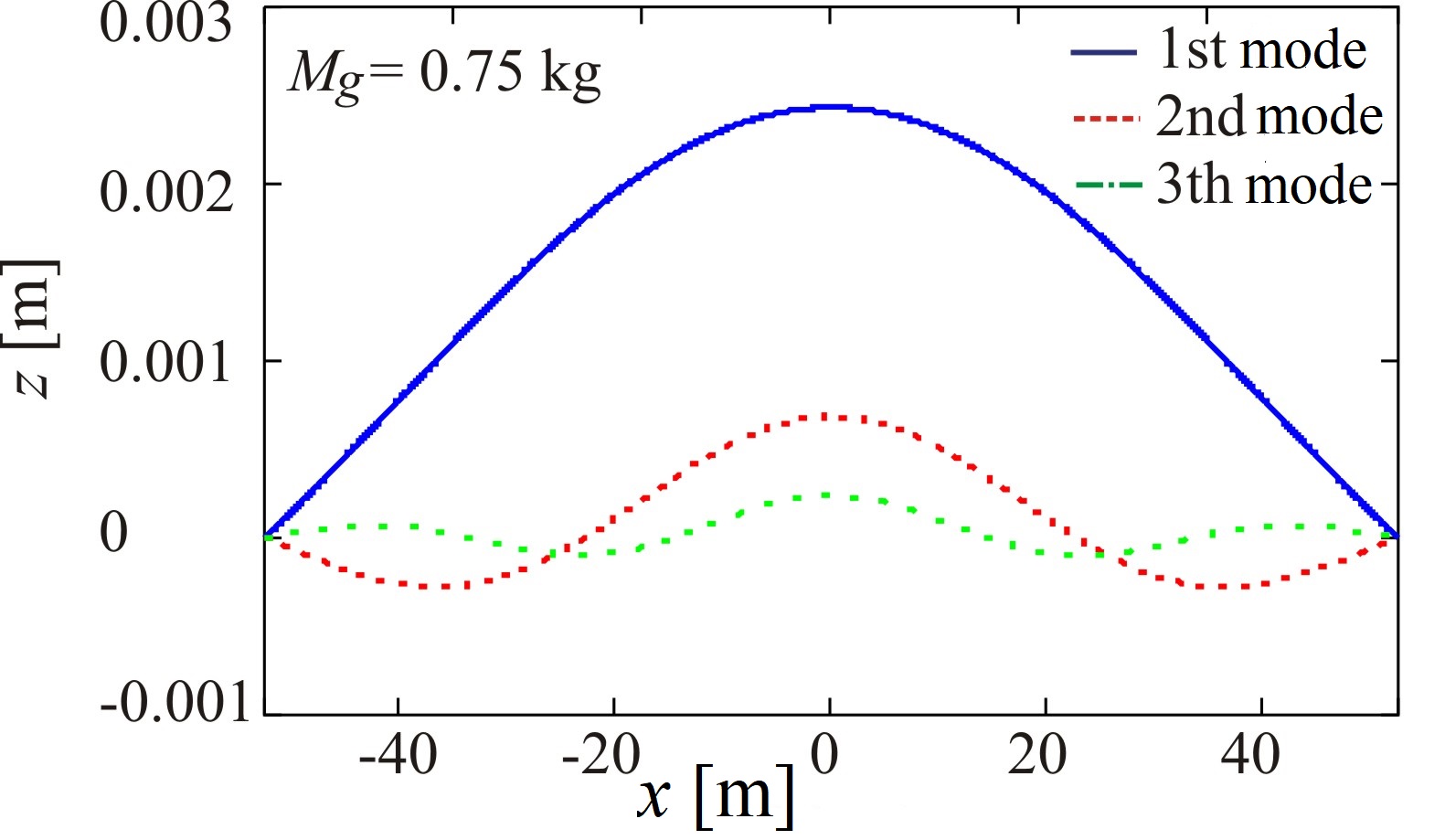}
\par\end{centering}
\caption{(Color online) Dynamics of the vibrating membrane. Maximum membrane deviations for the first (0,1), second  (0,2) and third (0,3) modes of
normal oscillations. % for the gas mass $M_{g} = 0.75$ kg in the toroidal shell.
}
\label{Vibration}
\end{figure}
In summary, there are two limits for the radii of the sail: i. the lower limit is due to the sail stability requirement demanding the stresses to be within limits of elasticity; ii. the upper limit is due to requirement of the successful deployment. The radius of the sail has to simultaneously satisfy both requirements being greater than the lower limit and less than the upper limit. This means that the lower limit has to be less than the upper limit. The dependence of the limits of the toroidal shell-membrane radii on the gas mass for choice of $r = 0.1$ m is presented in Fig. \ref{Lim}. This is kind of a solar sail phase diagram. The coordinates of the crossover point are: $R = 54.5$ m and $M_{g} = 0.870$ kg. The sails with the radius and the gas mass within the area laying to the left of the crossover point and defined at the top by the upper limit and at the bottom by the lower limit curves are deployable and stable.
The sails with the radius and gas mass to the right of the crossover point are either unstable and/or not deployable. For example, in the case of CP1 Polyimide film the sail membrane of $R = 51$ m attached to the toroidal shell of $r = 0.1$ m inflated by 0.75 kg of the hydrogen is the optimal fit.
The same kind of the diagram can be calculated for other choices of toroidal shell radius $r$. The location shift of the crossover point with the variation of $r$ from 0.1 m to 0.25 m is quit small, within 1.5 m and 60 g.

\iffalse
The calculated $a_{c}/g$ ratio presented in Fig. \ref{Ratio} exceeds
this estimate for the certain range of toroidal shell radius $r$ values for a choice
of sail radius  $R<$ 55 m. For example, the ratio is well above the estimate
for $R=45$ m as $r>0.1$ m. The same time the ratio is above the estimate for the gas when $R<60$ m. This means
that for the certain choice of sail size the less value of toroidal shell radius is
preferable as far as the $a_{c}/g$ ratio allows the sail to be deployed and
the toroidal shell and the sail membrane stay within the limit of elasticity. In the case of CP1 Polyimide films the sail membrane of $R=57$ m attached to the toroidal shell with cross-sectional radius $r=0.1$ inflated by 1 kg of the  hydrogen is the best fit that satisfies the condition to be within the limit of elasticity and be deployed.

\fi

For a sail with $R = 51$ m and inflatable toroidal shell with $r = 0.1$ m, $M_{g} = 0.750$ kg of the gas in the toroidal shell and for the tension of the membrane given by Eq. (\ref{Fm}) we calculated the first three normal transverse vibration modes of the deployed and stretched  circular sail. For this tension and the corresponding geometric characteristics of the membrane, the maximum deviation of the solar sail membrane was found. %for the first three forms of normal oscillations.
The results of these calculations are shown in Fig. \ref{Vibration}. Each graph corresponds to the instant of time when the maximum deviation of the membrane from its equilibrium position is reached. Our results show that the total deviations of the membrane from the equilibrium position with free modes of oscillations will not exceed a few millimeters and the maximal deviation is observed for the first mode. However, one should mention that these vibrations are related to the static deployed sail and only are transverse or out of plane without taking into account any damping in particular the dumping due to radiation pressure. These results could dramatically change when one considers the vibration during the dynamics of opening the sail when the gas is inserting into the toroidal shell. This aspect along with the in-plane vibrations and dumping is out of the scope of this paper and will be considered somewhere later.

\section{Conclusions}
\label{conclusions}
In this article we present the theoretical study of the circular solar sail
with inflatable torus-shaped shell. In the framework of a strict mathematical approach based on classical theory
of elasticity we justify an idea of the deployment and stretching of the
solar sail membrane attached to the inflatable toroidal shell.
We obtained the analytical expressions that
can be applied to a wide range of materials for both the
shell and sail membrane. The presented numerical example demonstrates the ability of the developed theory to provide sound estimates of the stresses of the inflatable shell and the solar sail membrane solely based on the geometry, gas mass and elastic properties of the materials. Analysis of these expressions shows the strong dependence of stresses and strains on the membrane
radius and on the pressure of the filled gas, but negligible dependence on the radius of the toroidal shell. We performed numerical calculations for the
sail of radius of 30 to 100 m made of CP1
membrane and attached to the toroidal shell of the same material with the varied cross-section
radius and gas mass. We predict
that by introducing the gas into the inflatable toroidal rim one can deploy and stretch circular solar sail membrane up to 54 m of radius. It is worth noting that the same calculations
%for the dependencies of stresses on the toroidal shell and solar sail membrane
can be easily performed  for the other
suggested materials for a solar sail such as Kapton or aluminized Mylar or others \cite{MatlKez2008}. It is shown that the normal out of plane vibrations
of the sail membrane under tension caused by gas pressure are
negligible.

To conclude, our theoretical approach and calculations demonstrate the feasibility of deployment and stretching of the solar sail constructed as a thin circular membrane attached to the inflatable toroidal shell.
At this point, an experimental research is necessary to find out the level of feasibility of such a solar sail for future exploration of the solar system and beyond by mean of the electromagnetic pressure propellant.


\begin{thebibliography}{99}
\bibitem{pol} E. N. Polyakhova, Kosmicheskii Polet s Solnechnim Parusom
(Space Solar Sailing) (\textit{in Russian) }Moscow, Nauka, 1986.

\bibitem{Colin} C. R. McInnes, Solar Sailing - Technology, Dynamics and
Mission Applications. Springer, Praxis Publishing, Chichester, 1999.

\bibitem{Matloff3} G. L. Matloff, Deep Space Probes: To the Outer Solar
System and Beyond. Springer/Praxis Books, 2005.

\bibitem{MatloffValpetti} G. Vulpetti, L. Johnson, G. L. Matloff, Solar
Sails - A Novel Approach to Interplanetary Travel. Copernicus Books, 2008.

\bibitem{9} J. M. Fernandez, V. J. Lappas, A. J. Daton-Lovett, Completely
stripped solar sail concept using bi-stable reeled composite booms. Acta
Astronautica \textbf{69}, 78--85 (2011).

\bibitem{Genta1999} G. Genta and E. Brusa, The parachute sail with
hydrostatic beam: a new concept for solar sailing. Acta Astronautica \textbf{%
44}, 133--140 (1999).

\bibitem{Deployment4} M. Salama, C. White, and R. Leland, Ground
demonstration of a spinning solar sail deployment concept. J. Spacecraft
Rock. \textbf{40}, 9--14 (2003) https://doi.org/10.2514/2.3933.

\bibitem{Dep4} J. M. Fernandez, Advanced deployable shell-based composite booms for small satellite structural applications including solar sails, Corpus ID: 114881473, Engineering (2017).

\bibitem{Dep5}  J. M. Fernandez, G. K. Rose, C. J. Younger, G. D. Dean,  J. E. Warren, O. R. Stohlman, and W. Wilkie, NASA’s advanced solar sail propulsion system for low-cost deep space exploration and science missions that uses high performance rollable composite booms, Corpus ID: 113546105, Physics (2017).


\bibitem{8} M. B. Quadrelli and J. West, Sensitivity studies of the
seployment of a square inflatable solar sail with vanes,\ Acta Astronautica
\textbf{65}, 1007--1027 (2009).

\bibitem{91}J. M. Fernandez, L. Visagie, M. Schenk, O. R. Stohlman, G. S. Aglietti, V. J. Lappas, and S. Erb, Design and development of a gossamer sail system for deorbiting in low earth orbit. Acta Astronautica |textbf{103}, 204--225 (2014). doi:10.1016/J.ACTAASTRO.2014.06.018

\bibitem{MemorySail} A. Boschetto, L. Bottini, G. Costanza, and M. E. Tata,
Shape memory activated self-deployable solar sails: small-scale prototypes
manufacturing and planarity analysis by 3D laser scanner, Actuators \textbf{8%
}, 38 (2019).

\bibitem{MemorySail2} A. Boschetto, L. Bottini, G. Costanza, and M. E. Tata,
A novel self-deployable solar sail system activated by shape memory alloys,
Aerospace \textbf{6}, 78 (2019).

\bibitem{Deployment1} B. Vatankhahghadim and C. J. Damaren, Solar sail
deployment dynamics, Adv. Space Res. \textbf{67}, 2746--2756 (2021).

\bibitem{Deployment2} V. Parque, W. Suzaki, S. Miura, A.Torisaka, T.
Miyashita, and M. Natori, Packaging of thick membranes using a multi-spiral
folding approach: Flat and curved surfaces, Adv. Space Res. \textbf{67}, 2589--2612 (2021).

\bibitem{Deployment3} L. T. Hibbert and H. W. Jordaan, Considerations in the
design and deployment of flexible booms for a solar sail, Adv. Space Res.
Adv. Space Res. \textbf{67}, 2716--2726 (2021).

\bibitem{VKRK20221} V. Ya. Kezerashvili and R. Ya. Kezerashvili, On deployment of solar sail with superconducting current-carring wire, Acta Astronautica \textbf{189}, 196--198 (2021).

\bibitem{VKRK20222} V. Ya. Kezerashvili and R. Ya. Kezerashvili, Solar sail with superconducting circular current-carrying wire, Adv. Space Res.
Adv. Space Res. \textbf{69}, 664--676 (2022).

\bibitem{5} G. F. Pezditz, Erectanle space structures-ECHO Satellites,\ NASA
N62-12545, 1962.

\bibitem{4} D. Cadogan, J. Stein, and M. Grahne, Inflatable composite
habitat structures for lunar and mars exploration, Acta Astronautica,
\textbf{44}, 399--406 (1999).

\bibitem{Strobl1} J. Strobl, The hollow-body solar sail, JBIS, \textbf{42},
515--520 (1989).

\bibitem{Hayn} D. Hayn, The orbital torus solar sail vehicle (ORTOSS:
Orbitaler Torus Sonnensegler), Luft und Raumfahrt, \textbf{11}, 34--36
(1990).

\bibitem{6} R. E. Freeland, G. D. Bilyeu, G. R. Veal, M. D. Steiner, and D.
E. Carson, Large Inflatable Deployable Antenna Flight Experiment Results,\
Acta Astronautica, \textbf{41}, 267-277 (1997).

\bibitem{Strobl2} J. Strobl, The hollow-body solar sail as a transporter of
a Radio Telescope, JBIS \textbf{47}, 67--70 (1994).

\bibitem{Leigh} L. M. Leigh and M. L. Tinker, Dynamic characterization of an
inflatable concentrator for solar thermal propulsion, J. Spacecraft and
Rocket, \textbf{40}, 24--27 (2003).

\bibitem{Tinker} K. B. Smalley, M. L. Tinker, and W. S. Taylor, Structural
modeling of a five-meter thin-film inflatable antenna/concentrator, J.
Spacecraft and Rocket, \textbf{40}, 27--29 (2003).

\bibitem{Matloff1} G. L. Matloff, The beryllium hollow-body solar sail and
interstellar travel, JBIS \textbf{59} 349--354 (2006).

\bibitem{Kezmetloff1} R. Ya. Kezerashvili and G. L. Matloff, Solar radiation
and the beryllium hollow-body sail: 1. The ionization and disintegration
effects, JBIS \textbf{60}, 169--179 (2007).

\bibitem{Kezmetloff2} R. Ya. Kezerashvili and G. L. Matloff, Solar radiation
and the beryllium hollow-body sail: 2. Diffusion, recombination and erosion
processes, JBIS \textbf{61} 47-57 (2008).

\bibitem{KezASR2021} R. Ya. Kezerashvili, O. L. Starinova, A. S. Chekashov,
and D. J. Slocki, A torus-shaped solar sail accelerated via thermal
desorption of coating, Adv. Space Res. \textbf{67}, 2577--2588  (2021).

\bibitem{Timoshenko51} S. Timoshenko and J. N. Goodier, Theory of
Elasticity, 2$^{nd}$ Edition, McGraw-Hill Book Com., New York, USA, 1951.

\bibitem{Landau7} L.~D. Landau and E.~M. Lifshitz, Theory of Elasticity, 3$%
^{rd}$ English Edition, Revised and Enlarged, Pergamon Press, Oxford, UK,
1986.

\bibitem{Kraus1967} H. Kraus, Thin Elastic Shells. Wiley, New York 1967.

\bibitem{Handbook} A. D. Polyanin and V. F. Zaitsev, Handbook of Ordinary
Differential Equations, CRC\ Press, Taylor \& Francis, Boca Raton, USA,
2018.

\bibitem{CP1} A. Peloni, D. Barbera, S. Laurenzi, C. Circi, Dynamic and
structural performances of a new sailcraft concept for interplanetary
missions, Scientific World Journal, Volume 2015, Article ID 714371, 14 pages
http://dx.doi.org/10.1155/2015/714371.

\bibitem{LesJ} Les Johnson, Privet communication.

\bibitem{MatlKez2008}G. L. Matloff and R. Ya. Kezerashvili, Interstellar solar sailing: a figure of merit for monolayer sail, JBIS \textbf{61}, 330--333 (2008).



% \bibitem{} \textbullet \qquad Johnson, L., Whorton, M., Heaton, A., Pinson, R., Laue, G. and Adams, C., 2011. NanoSail-D: A solar sail demonstration mission. Acta Astronautica 68, 571--575.

%\bibitem{} \textbullet \qquad Polyakhova, E. N., Ovchinnikov, M. Yu., and Tikhonov, A. A., 2018. To 130-th Birthday Anniversary of Friedrich Tsander (1887-1933): Ten New Russian Books in Astrodynamics as the Honorable Contribution to his Memory, AIP Conf. Proc. 1959, 040015 (2018).

%\bibitem{} Zubrin, R. 1995, in ASP Conf. Ser.\U{f0a0}74, Progress in the Search for

%\bibitem{} Extraterrestrial Life, ed. G. S. Shostak (San Francisco, CA: ASP), 487

%\bibitem{} Zubrin, R. 2011, The Case for Mars: The Plan to Settle the Red Planet and

%\bibitem{} Why We Must (New York: Free Press)

%\bibitem{} Zubrin, R. 2019, The Case for Space: How the Revolution in Spaceflight Opens

%\bibitem{} Up a Future of Limitless Possibility (Buffalo, NY: Prometheus Books)

%\bibitem{} Zubrin, R. M., \& Andrews, D. G. 1991, JSpRo, 28, 197
\end{thebibliography}
\end{document}